\documentclass[a4paper,11pt]{article}
\usepackage{pos}
\usepackage{float}

\def\faz{f(x_e)\,(1 - (\alpha Z)^2/2)}
\renewcommand{\logo}{\relax}

\usepackage{amsmath}
\def\dfrac{\frac}

\title{Radiative Corrections in Bound States: Recent Results}
\ShortTitle{Radiative Corrections in Bound States}

\author*[a]{Andrzej Czarnecki}
\author[a]{Artem O.~Davydov}

\affiliation[a]{Department of Physics, University of Alberta,\\
 T6G 2G7, Edmonton, Canada}

\emailAdd{andrzejc@ualberta.ca}
\emailAdd{davydov@ualberta.ca}

\abstract{ Two recent studies of radiative corrections to bound state
  properties are discussed. The change of the decay rate of a muon
  bound to a light nucleus has been calculated for several light
  nuclei $4\leq Z \leq 9$ with high precision, resolving a
  long-standing discrepancy between analytical and numerical results
  for oxygen ($Z=8$). The decay of parapositronium into three photons
  has been calculated including effects of the $Z$ boson. The
  resulting rate is many orders of magnitude smaller than previously
  estimated.}

\FullConference{Loops and Legs in Quantum Field Theory (LL2026)\\
12--17 April, 2026\\
Bayreuth, Germany\\}

\begin{document}
\maketitle

\section{Introduction}

The muon is similar to the electron in most respects: both are
leptons, carry identical electric charge, and have the same
spin. However, the muon is approximately 207 times heavier than the
electron. As a result of energy-momentum conservation, the muon can
decay into an electron, while the electron is stable. The decay rate
of a free muon is given by
\begin{equation}
    \Gamma_0=\dfrac{G_F^2m^5_\mu}{192\pi^3},
\end{equation}
where $m_\mu\approx 105.65\text{ MeV}$ is the muon mass and $G_F$ is
the Fermi constant. In the present discussion we focus on the ratio of
the decay rate $\Gamma$ of a muon bound to a light nucleus to
$\Gamma_0$ of a free muon. Muonic atoms offer probes of physics beyond
the Standard Model~\cite{Uesaka:2024tfn,Matias:2025vrq}, especially
for searches for the charged-lepton-flavor-violating (CLFV) coherent
muon-to-electron conversion,
$\mu^- N \to e^-
N$~\cite{Miscetti:2025uxk,COMET:2025sdw,Nishiguchi:2025bsr,Yamamoto:2023pxc}.
For those experiments, the ordinary muon decay
$\mu^- N \longrightarrow e^-N + \bar{\nu}_e + \nu_\mu$, which does not
violate lepton number conservation, is an irreducible background.

For the muon decay in orbit (DIO), an analytical perturbative calculation
accurate to order $(\alpha Z)^2$ was carried out in Ref.~\cite{Uberall:1960zz}:
\begin{equation}
   \Gamma/\Gamma_0 = f(x_e)\bigg(1 - \frac{(\alpha Z)^2}{2}\bigg)
   + \mathcal{O}((\alpha Z)^3),
   \label{eq:Uberall-smallZ}
\end{equation}
where $f(x_e)$ accounts for the finite electron mass,
\begin{equation}
    f(x_e) = 1 - 8x_e^2 - 24x_e^4\ln x_e + 8x_e^6 - x_e^8 = 0.99981,
    \quad
    x_e \equiv \frac{m_e}{m_\mu} = 4.8363 \times 10^{-3}.
\end{equation}
However, a fully relativistic calculation of the total DIO rate based
on a partial-wave expansion~\cite{Watanabe:1993emp} showed a
significant deviation from this prediction for oxygen ($Z=8$). The
numerical result reported in Ref.~\cite{Watanabe:1993emp} is
$\Gamma/\Gamma_0 = 0.994$, while formula~\eqref{eq:Uberall-smallZ}
gives $\Gamma/\Gamma_0=0.9981$.  The expected next-to-leading
$(\alpha Z)^3$ correction, $(\alpha Z)^3 \sim 2\times10^{-4}$, is more
than an order of magnitude smaller than the observed discrepancy and
cannot explain it. In \cite{Czarnecki:2025phw} we reexamined the DIO
rate for oxygen and showed that the discrepancy originated from too
early truncation of a slowly convergent partial-wave series. To
support this conclusion, we also calculated the decay rate for other
nuclei in the range $4\leq Z\leq 9$ and showed that the deviation
between our all-order-in-$\alpha Z$ numerical results and the
analytical expression~\eqref{eq:Uberall-smallZ} decreases smoothly as
$Z\longrightarrow0$, confirming that the difference is accounted for
by higher-order corrections $\mathcal{O}((\alpha Z)^3)$.

Below we describe that work, and then a second bound-state decay where
explicit calculation likewise corrects an earlier estimate:
parapositronium into three photons. It is forbidden in pure QED but
opened by the C-violating weak interaction. Its rate proves orders of
magnitude below the original guess.

\section{Theoretical framework}

Following Refs.~\cite{gilinsky60,Aslam:2020pqn}, we factorize the
bound decay into the cascade $(Z\mu^-) \to [Ze^-]A$ and
$A \to \bar{\nu}_e \nu_\mu$, with $A$ a fictitious massive neutral
vector carrying the flavor of the neutrino pair. Here $(Z\mu^-)$ is a
muon bound in the nuclear field and $[Ze^-]$ a continuum electron in
that field. The decay rate reads
\begin{equation}
\label{eq:intermediate-boson}
\Gamma\!\left((Z\mu^-) \to [Ze^-]\,\bar{\nu}_e \nu_\mu\right)
=\frac{256\pi\,\Gamma_0}{g^{2} m_\mu}\int_{0}^{z_{\max}}\! dz\;
\Gamma\!\left((Z\mu^-) \to [Ze^-]A\right)\, z^{3},
\end{equation}
where the dimensionless variable $z$ is defined as
\begin{equation}
  z \equiv \frac{m_A}{m_\mu}, \qquad
  0 \le z \le z_{\max} = \frac{E_\mu-E_e}{m_\mu}\,,
\end{equation}
and $g$ is the weak-coupling constant, $m_A$ is the invariant mass of the
intermediate boson~$A$, with $E_\mu$ and $E_e$ being the total energies of
the initial-state muon and the final-state electron, respectively.

The two-body decay rate $\Gamma((Z\mu^-)\to [Ze^-]A)$ is given by
\begin{equation}
\label{eq:Gamma-intermediate}
\Gamma\!\left((Z\mu^-) \to [Ze^-]A\right)
=\frac{1}{64\pi^5}\!\int\!\frac{d^{3}\vec{p}_e}{p_eE_e}\;
      \frac{d^{3}\vec{k}}{k_{0}}\;
      \delta\!\left(E_\mu-E_e-k_{0}\right)
      \bigl|{\cal M}_{(Z\mu^-)\to [Ze^-]A}\bigr|^{2},
\end{equation}
where $p_e^\nu=(E_e,\vec{p}_e)$ with $p_e\equiv|\vec{p}_e|$ and
$k^\nu=(k_0,\vec{k})$ are the four-momenta of the outgoing electron and
the intermediate boson $A$, respectively. The squared amplitude reads
\begin{equation}
\label{current}
\bigl|{\cal M}_{(Z\mu^-)\to [Ze^-]A}\bigr|^{2}
=
\frac{1}{2}
\sum_{\mu_\mu}\sum_{\kappa_e\mu_e}
\frac{g^{2}}{2}\,
J^{\alpha}(\vec{k})\,
J^{\dagger\beta}(\vec{k})\,
\left(-g_{\alpha\beta}+\frac{k_{\alpha}k_{\beta}}{m_A^{2}}\right),
\end{equation}
where the leptonic currents $J^{\alpha}(\vec{k})$ are overlap integrals
between the muon wave function $\Psi_{\mu_\mu}(\vec{r})$ and the electron
wave function $\psi_{E_e\kappa_e\mu_e}(\vec{r})$:
\begin{equation}
\label{eq:current muon-electron}
J_{\alpha}(\vec{q})
=\int d^{3}\vec{r}\,
e^{-i\vec{q}\cdot\vec{r}}\,
\psi^{\dagger}_{E_e\kappa_e\mu_e}(\vec{r})\,\gamma_{0}\gamma_{\alpha}P_{L}\,
\Psi_{\mu_\mu}(\vec{r}).
\end{equation}
The angular integrations in~(\ref{eq:current muon-electron})
and~(\ref{current}), as well as the summations over the total angular
momentum projections $\mu_{\mu}$ and $\mu_e$, can be performed
analytically~\cite{Czarnecki:2011mx, Watanabe:1987su,
  Watanabe:1993emp, Kaygorodov:2025yag}, leading to the following
representation of the decay rate:
\begin{equation}
    \frac{\Gamma}{\Gamma_0} = \sum_{\kappa_e}\frac{\Gamma_{\kappa_e}}{\Gamma_0}\,,
    \label{partial wave}
\end{equation}
where $\Gamma_{\kappa_e}$ are the partial-wave contributions evaluated
numerically.

\section{Numerical results and discussion}\label{sec:numerics}

We evaluate $\Gamma/\Gamma_0$ from
Eqs.~\eqref{eq:intermediate-boson}--\eqref{partial wave} for
point-Coulomb nuclei, $4 \leq Z \leq 9$.  The muon enters through the
closed-form Dirac–Coulomb ground state \cite{rosebook}; the continuum
electron comes from integrating the Dirac equation with
\textsc{RADIAL} \cite{salvat1995radial,salvat2019radial} in
double-precision.  The radial integrals oscillate strongly and are
handled with the adaptive \texttt{DQAGS} routine of
\textsc{QUADPACK}~\cite{piessens2012quadpack} at $10^{-9}$ tolerance.
We extend the $\kappa_e$ sum until term-to-term changes reach
$10^{-5}$--$10^{-6}$ and fix the tail with a geometric estimate.

In Table~\ref{tab:ratio} we present the converged results for
$\Gamma/\Gamma_0$ and compare them with
Refs.~\cite{Uberall:1960zz,Watanabe:1993emp,Uesaka:2026tdd}.  In
Ref.~\cite{Uesaka:2026tdd} the charge distribution based on modern
experimental data was taken into account and the decay rates were
averaged over different isotopes of each element. The agreement
between Ref.~\cite{Uesaka:2026tdd} and our results confirms that the
effect of the finite nuclear charge distribution is negligible for
light nuclei.

\begin{table}[t]
\centering
\caption{The total bound-muon decay rate normalized to the free-muon
  rate, $\Gamma/\Gamma_{0}$, for various atomic numbers $Z$.  The
  second and third columns show results of the fully relativistic
  calculation of the present work (PW) and
  Ref.~\cite{Watanabe:1993emp}, respectively.  The fourth column shows
  results from the perturbative $\alpha Z$ expansion of
  Ref.~\cite{Uberall:1960zz} corrected for the finite electron mass.
  The fifth column shows isotope-averaged results from
  Ref.~\cite{Uesaka:2026tdd}.  The sixth column, $\Delta(Z)$, shows
  the difference between the $\alpha Z$ results and the present work
  in units of $10^{-3}$.}
\label{tab:ratio}
\begin{tabular}{cccccc}
  $Z$ & PW & Ref.~\cite{Watanabe:1993emp} & Ref.~\cite{Uberall:1960zz}
      & Ref.~\cite{Uesaka:2026tdd} & $\Delta(Z)\,[10^{-3}]$ \\
\hline
 4 & 0.9993 &       & 0.9994 &        & 0.045 \\
 5 & 0.9990 &       & 0.9991 &        & 0.099 \\
 6 & 0.9987 &       & 0.9989 & 0.9987 & 0.188 \\
 7 & 0.9982 &       & 0.9985 & 0.9982 & 0.322 \\
 8 & 0.9976 & 0.994 & 0.9981 & 0.9977 & 0.506 \\
 9 & 0.9969 &       & 0.9977 & 0.9970 & 0.757 \\
\end{tabular}
\end{table}

\begin{figure}[htb]
\centering
\input{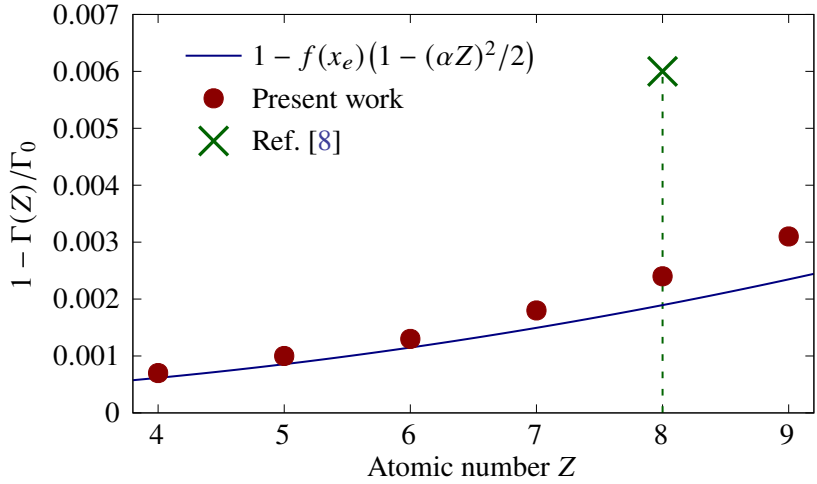}
\caption{The quantity $1-\Gamma/\Gamma_0$ as a function of $Z$.
Red circles: present work. Blue solid line: analytical approximation
$1-\faz$, where $f(x_e)$ accounts for the finite electron mass.
Green cross: result of Ref.~\cite{Watanabe:1993emp} for $Z=8$.
The systematic excess of the red circles above the solid line reflects
higher-order corrections in $\alpha Z$.}
\label{fig:uberall}
\end{figure}

Our results are visualized in
Fig.~\ref{fig:uberall}, which plots $1 - \Gamma/\Gamma_0$ versus $Z$.
The solid line represents the $(\alpha Z)^2$ prediction
$1 - f(x_e)(1 - (\alpha Z)^2/2)$ of Ref.~\cite{Uberall:1960zz}, while the
red circles denote the present all-order-in-$\alpha Z$ values. Our
results lie consistently above the $(\alpha Z)^2$ curve, and
the magnitude of this excess grows with $Z$, as quantified by the positive
$\Delta(Z)$ values in Table~\ref{tab:ratio}. This trend is consistent with
a negative next-order correction to $\Gamma/\Gamma_0$ in the $\alpha Z$
expansion. In contrast, the result of Ref.~\cite{Watanabe:1993emp} for
$Z=8$ (green cross in Fig.~\ref{fig:uberall}) deviates significantly from
both our data and the $(\alpha Z)^2$ prediction.

The origin of the discrepancy is revealed by Table~\ref{Z=8}, which
contains the partial sums
$\sum_{|\kappa_e|\le\kappa_{\max}}\Gamma_{\kappa_e}/\Gamma_0$ for
different truncation limits $\kappa_{\max}$. The series converges
slowly: even at $\kappa_{\max}=29$ the result
$\Gamma/\Gamma_0 = 0.99389$ is still far from the converged
value. Ref.~\cite{Watanabe:1993emp} terminated the expansion at
$L=31$, obtaining $\Gamma/\Gamma_0 = 0.994$, in good agreement with
our unconverged partial sum at comparable $\kappa_{\max}=29$. Only by
extending the summation up to $\kappa_{\max}=59$ and extrapolating the
remaining tail allowed us to obtain the converged result
$\Gamma/\Gamma_0 = 0.99760$, which is consistent with the analytical
prediction~\eqref{eq:Uberall-smallZ}.

\begin{table}[h]
\centering
\caption{Convergence of the partial-wave expansion for
${}^{16}\mathrm{O}$ ($Z=8$). Shown are the partial sums
$\sum_{|\kappa_e|\le\kappa_{\max}}\Gamma_{\kappa_e}/\Gamma_0$
as a function of the cutoff $\kappa_{\max}$.}
\label{Z=8}
\begin{tabular}{ccc}
\hline
$\kappa_{\max}$ & Present work & Ref.~\cite{Watanabe:1993emp} \\
\hline
25 & 0.98871 & \\
29 & 0.99389 & 0.994 \\
35 & 0.99661 & \\
40 & 0.99728 & \\
45 & 0.99750 & \\
50 & 0.99757 & \\
55 & 0.99759 & \\
59 & 0.99760 & \\
\hline
\end{tabular}
\end{table}

\section{Decay of parapositronium into three photons}
Positronium, with two spin-1/2 components:  an electron and a
positron, has a spin-0 ground state called parapositronium (pPs) and
a spin-1 slightly excited state called orthopositronium (oPs). By the
Landau-Yang theorem, oPs cannot decay into two photons, because of
Bose-Einstein statistics and angular momentum conservation. A decay of
oPs results in at least three photons. On the other hand, pPs does
decay into two photons; can it decay also into three? The role of
symmetries in the decays of positronium and related systems such as the Ps$^-$ 
ion has very recently been discussed in \cite{Sani:2026onr},
where further references can be found (see also \cite{Bass:2019ibo}).

It turns out that in pure quantum electrodynamics (QED) pPs decays
only into even numbers of photons because of charge-conjugation
symmetry C. However, weak interactions violate C and enable the decay
$\text{pPs}\to 3\gamma$.  The rate of this decay was first estimated in
\cite{Bernreuther:1981ah}. Ref.~\cite{Pokraka:2017ore} calculated the
contribution of $W$ loops to this channel. More recently, the full
one-loop amplitude, including $Z$-boson contributions, has been determined
\cite{Czarnecki:2026ect}. 

\begin{figure}[htb]
\centering
\includegraphics[width=0.98\linewidth]{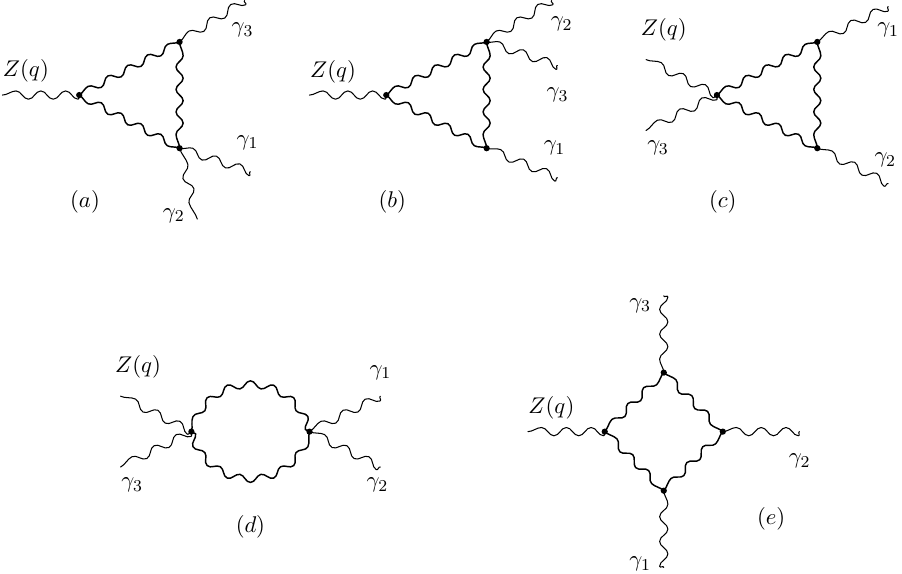}
\caption{Representative $W$-loop diagrams contributing to
  $Z^\star(q)\to\gamma(k_1)\gamma(k_2)\gamma(k_3)$. Diagrams $(a)$ and
  $(b)$ contain one $AWW$ and one $AAWW$ photon vertex, $(c)$ contains
  one $ZAWW$ and two $AWW$ vertices, $(d)$ contains one $ZAWW$ and one
  $AAWW$ vertex, and $(e)$ contains three $AWW$ vertices. The photon
  permutations are generated by $\mathcal{S}_3$.}
\label{fig:zw-topologies}
\end{figure}

These explicit calculations showed that the rate of
$\text{pPs}\to 3\gamma$ is smaller by many orders of magnitude than
first estimated. One reason for this suppression is the vanishing of
some classes of diagrams \cite{Czarnecki:2026ect}. Here we illustrate
that vanishing with the example of pPs annihilation that produces only
a virtual $Z$ boson, which then decays into three photons through a
$W$-loop, as shown in Fig.~\ref{fig:zw-topologies}.

A $Z$ boson can decay into $3\gamma$ \cite{Glover:1993nv}. That it
cannot mediate $\text{pPs} \to 3\gamma$ is best seen in the unitary
gauge. Consider first the amplitude of annihilation $\text{pPs} \to
Z^\star$. Since pPs has no spin and $Z$ has spin 1, that amplitude
transforms as a Lorentz vector under rotations and boosts, so it must
be proportional to $q$, the 4-momentum of pPs, the only 4-vector
relevant for this process. The structure of the total amplitude is
\begin{equation}
  \label{eq:1}
  {\mathcal M}\left(\text{pPs}\to Z^\star \to 3\gamma\right) \propto
  q^\mu D^Z_{\mu\nu}  {\mathcal M}^\nu\left(Z^\star \to 3\gamma\right),
\end{equation}
where we denote by $D^a_{\mu\nu}$ propagators of vector bosons, $a=W,Z$.
In the unitary gauge we use
\begin{equation}
 D^a_{\alpha\beta}(p)
 =\frac{i}{M_a^2-p^2}
 \left(g_{\alpha\beta}-\frac{p_\alpha p_\beta}{M_a^2}\right),
 \quad
 D^{-1}_{\alpha\beta}(p)
 =i\left[(p^2-M_a^2)g_{\alpha\beta}-p_\alpha p_\beta\right],\quad a=W,Z.
 \label{eq:zw-prop}
\end{equation}
The amplitude in Eq.~\eqref{eq:1} becomes
\begin{equation}
  \label{eq:2}
    {\mathcal M}\left(\text{pPs}\to Z^\star \to 3\gamma\right) \propto
  q_\nu {\mathcal M}^\nu\left(Z^\star \to 3\gamma\right).
\end{equation}
Below we show the vanishing of this contraction of the amplitude
$ {\mathcal M}^\nu\left(Z^\star \to 3\gamma\right)$ with the incoming
momentum $q$ of the $Z$-boson.

\begin{figure}[htb]
\centering
\includegraphics{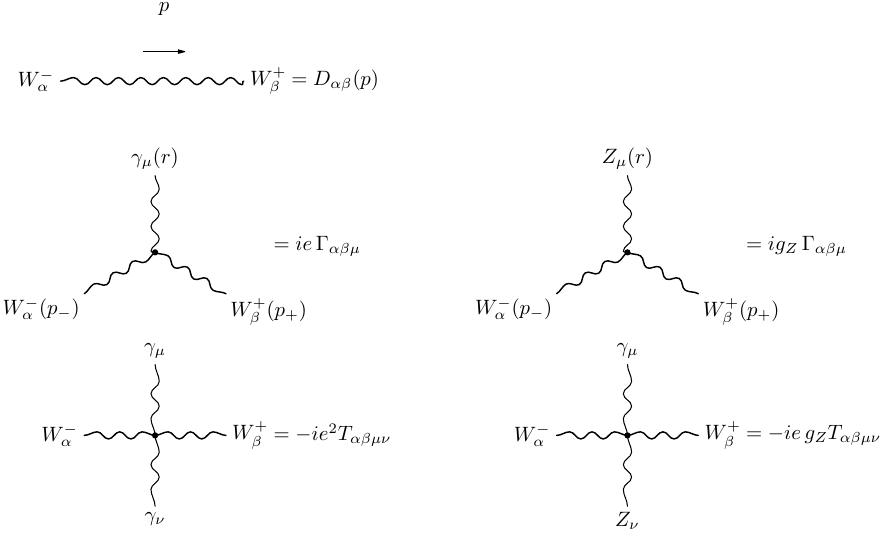}
\caption{Unitary-gauge propagator and vertices used in the calculation.
All momenta entering the vertex expressions are taken as incoming.}
\label{fig:zw-rules}
\end{figure}

With all momenta at a vertex taken as incoming, define
\begin{align}
 \Gamma_{\alpha\beta\mu}(p_-,p_+,r)
 &=(p_--p_+)_\mu g_{\alpha\beta}
 +(p_+-r)_\alpha g_{\beta\mu}
 +(r-p_-)_\beta g_{\mu\alpha},
 \label{eq:zw-gamma}\\
 T_{\alpha\beta\mu\nu}
 &=2g_{\alpha\beta}g_{\mu\nu}
   -g_{\alpha\mu}g_{\beta\nu}
   -g_{\alpha\nu}g_{\beta\mu}.
 \label{eq:zw-tensor}
\end{align}
Writing $g_Z=e c_W/s_W$, the required Feynman rules \cite{Denner:2019vbn} are (we use $e>0$)
\begin{align}
 V^A_{\alpha\beta\mu}
 &=ie\,\Gamma_{\alpha\beta\mu},
&
 V^Z_{\alpha\beta\mu}
 &=ig_Z\,\Gamma_{\alpha\beta\mu},
 \label{eq:zw-fr-cubic}\\
 V^{AA}_{\alpha\beta\mu\nu}
 &=-ie^2T_{\alpha\beta\mu\nu},
&
 V^{ZA}_{\alpha\beta\rho\mu}
 &=-ie\,g_ZT_{\alpha\beta\rho\mu}.
 \label{eq:zw-fr-quartic}
\end{align}
The diagrammatic conventions corresponding to
Eqs.~\eqref{eq:zw-prop}, \eqref{eq:zw-fr-cubic}, and
\eqref{eq:zw-fr-quartic} are shown in Fig.~\ref{fig:zw-rules}.

The elementary Ward identities following from
Eqs.~\eqref{eq:zw-fr-cubic} and \eqref{eq:zw-fr-quartic} are
\begin{align}
 q^\rho V^Z_{\alpha\beta\rho}(p_-,p_+,q)
 &=g_Z\left[
 D^{-1}_{\alpha\beta}(p_+)-D^{-1}_{\alpha\beta}(p_-)
 \right],
 \label{eq:zw-ward-cubic}\\
 q^\rho V^{ZA}_{\alpha\beta\rho\mu}(p_-,p_+,q,k)
 &=g_Z\left[
 V^A_{\alpha\beta\mu}(p_-,p_++q,k)
 -V^A_{\alpha\beta\mu}(p_-+q,p_+,k)
 \right].
 \label{eq:zw-ward-quartic}
\end{align}
Placing Eq.~\eqref{eq:zw-ward-cubic} between the two adjacent propagators
and using Eq.~\eqref{eq:zw-prop} gives
\begin{equation}
 D(p_-)\,q^\rho V^Z_\rho\,D(p_+)
 =g_Z\left[D(p_-)-D(p_+)\right].
 \label{eq:zw-prop-collapse}
\end{equation}
Thus contraction of a $ZWW$ vertex removes either adjacent propagator,
whereas contraction of a $ZAWW$ vertex gives the difference of two $AWW$
insertions, as displayed in Eq.~\eqref{eq:zw-ward-quartic}.

The five representative topologies are shown in
Fig.~\ref{fig:zw-topologies}. After summing over the six permutations
of the photons, denoted by  $\mathcal{S}_3$, the amplitude is organized as
\begin{equation}
 \mathcal{M}^{\nu}
 =\mathcal{S}_3\left[
 \frac{1}{2}\left(
 \mathcal{M}_a+\mathcal{M}_b+\mathcal{M}_d
 \right)
 +\mathcal{M}_c+\mathcal{M}_e
 \right]^{\nu}.
 \label{eq:zw-amplitude}
\end{equation}
The factor $1/2$ removes the double counting caused by interchanging the
two photons attached to an $AAWW$ vertex.

The two terms in Eq.~\eqref{eq:zw-amplitude} are separately transverse.
To see this, first remove the $Z$ insertion. The diagrams $(a)$, $(b)$,
and $(d)$ arise from the same electromagnetic skeleton: a closed $W$ loop
with one $AWW$ vertex and one $AAWW$ vertex. Inserting the $Z$ on either
of its two $W$ propagators gives $(a)$ and $(b)$, while replacing its
$AWW$ vertex by a $ZAWW$ vertex gives $(d)$.
Equations~\eqref{eq:zw-ward-quartic} and
\eqref{eq:zw-prop-collapse} turn the contraction with $q$ into finite
differences between neighboring insertions. These differences telescope
around the closed loop, so
\begin{equation}
 q_\nu\,\mathcal{S}_3\,
\left(
 \mathcal{M}_a+\mathcal{M}_b+\mathcal{M}_d
 \right)^{\nu}=0.
 \label{eq:zw-abd-zero}
\end{equation}

The diagrams $(c)$ and $(e)$ arise from a second skeleton, containing
three $AWW$ vertices. Inserting the $Z$ on its three $W$ propagators
produces the box diagrams $(c)$, while replacing any $AWW$ vertex by a
$ZAWW$ vertex produces the triangle diagrams $(e)$. The same telescopic
cancellation gives
\begin{equation}
 q_\nu\,\mathcal{S}_3
 \left(
 \mathcal{M}_c+\mathcal{M}_e
 \right)^{\nu}=0.
 \label{eq:zw-ce-zero}
\end{equation}
The two sectors do not mix because contraction of the $Z$ leg preserves
the number of $AAWW$ vertices. Equations~\eqref{eq:zw-abd-zero} and
\eqref{eq:zw-ce-zero} therefore prove the vanishing of the complete
amplitude in Eq.~\eqref{eq:1}.

\section{Conclusion}\label{chap:Conclusion}

We have reviewed the fully relativistic calculation of the bound-muon decay
rate for light nuclei ($4\leq Z\leq 9$) that resolved the long-standing
discrepancy between the all-order-in-$\alpha Z$ numerical approach and the
perturbative analytical result of Ref.~\cite{Uberall:1960zz}. The
discrepancy was traced to a premature truncation of the slowly convergent
partial-wave series in Ref.~\cite{Watanabe:1993emp}. Comparison with
Ref.~\cite{Uesaka:2026tdd} confirms that finite nuclear charge distribution
effects are negligible for the light nuclei considered here.

We have also reviewed a calculation of $Z$-boson contributions to the
C-violating parapositronium decay into three photons. An early
estimate of its rate turned out to be much too large because some
diagrams are suppressed by small ratios of masses and some vanish
exactly. We provided an example of a vanishing contribution and
explained in some detail why it does not contribute.

\begin{acknowledgments}
  AC thanks Ting Gao for discussions of the amplitude of
  $\text{pPs}\to 3\gamma$ and of reasons why some classes of diagrams
  do not contribute.  This work was supported by the Natural Sciences
  and Engineering Research Council of Canada (NSERC).
\end{acknowledgments}


\end{document}